\documentstyle[aaspp4,flushrt]{article}     

\begin{document}

\vspace*{3.5cm}

\title{HI ``Tails'' from Cometary Globules in IC1396}

\author{G. H. Moriarty-Schieven}
\affil{Dominion Radio Astrophysical Observatory, 
National Research Council of Canada, Penticton, BC V2A 6K3, Canada.
(gms@jach.hawaii.edu)}
\authoraddr{Joint Astronomy Centre, 660 N. A'ohoku Pl., Hilo, HI  96720}

\author{T. Xie}
\affil{The Laboratory for Millimeterwave Astronomy,
Dept. of Astronomy, University of Maryland
College Park, MD  20742  (tao@astro.umd.edu)}

\and

\author{N. A. Patel}
\affil{Harvard-Smithsonian Center for Astrophysics
60 Garden St., Cambridge, MA  02138 (nimesh@smanap.harvard.edu)}

\begin{abstract}

IC 1396 is a relatively nearby (750 pc), large ($>$2\arcdeg), HII region 
ionized by a single O6.5V star and
containing bright-rimmed cometary globules.  We have made the first 
arcmin resolution 
images of atomic hydrogen toward IC 1396, and have found remarkable
``tail''-like structures associated with some of 
the globules and extending 
up to 6.5 pc radially away from the central ionizing star.  These HI
``tails'' may be material which has been ablated from the globule
through ionization and/or photodissociation and then accelerated away
from the globule by the stellar wind, but which has since drifted into the
``shadow'' of the globules.

This report presents the first results of the Galactic Plane Survey
Project recently begun by the Dominion Radio Astrophysical Observatory.

\end{abstract}

\keywords{ISM: globules -- ISM: HII Regions -- ISM: individual: IC 1396}

\section{Introduction}

IC 1396 (S131, Sharpless 1959) is a nearby (750 pc; Garrison \& Kormendy 1976),
large ($>$2\arcdeg), evolved HII region ionized by a single
O6.5V star HD 206267 located near its center.  Within the HII region are many 
dark globules, some with bright rims facing toward the central 
ionizing star (Pottasch 1956, 1958a,b; Osterbrock 1957).  Since the pioneering
work of Pottasch and Osterbrock, these globules have been studied in 
radio continuum, molecular emission, and optically (for a summary of these
studies see Patel et al. 1995).  For example, Patel et al. (1995) have
mapped the CO emission within the HII region, and have found that the 
bright-rimmed globules appear to trace a ring approximately 12.5 pc in 
radius which is expanding radially outward from the central star.  The 
expansion is apparently caused, not by the stellar wind or radiation 
pressure, but by a ``rocket effect'' (Harwit \& Schmid-Burgk 1983)
induced by the ionization of the 
inner-facing surface of the globules (the ``bright rims'').  Weikard et al. 
(1995) have also mapped the HII region in several transitions and isotopomers
of CO and in HI, but at moderate resolution (several arcmin).

In this report we present the first high-resolution ($\sim$1')
images of atomic hydrogen toward IC1396, 
showing remarkable ``tails'' of HI associated with some of the globules.  
(A more detailed analysis, presenting all of the data, awaits a later paper.)
These are the first results of the Galactic Plane Survey (GPS) recently 
begun by the Dominion Radio Astrophysical Observatory 
(DRAO)\footnote{The Dominion Radio Astrophysical Observatory is operated
as a national facility
by the Herzberg Institute for Astrophysics of the National Research Council
of Canada.}.    The GPS is being 
carried out by a consortium of Canadian and international astronomers, and
will provide an image of the Galactic Plane in the longitude interval 
$75^{\circ} - 145^{\circ}$ and latitude range $-3^\circ$ to $+5^\circ$, 
yielding an atomic hydrogen (HI) spectral line
data cube with 256 velocity channels and angular resolution of 
$1' \times 1' {\rm cosec}(\delta)$.  At the same time, continuum images at
1420 MHz and 408 MHz are obtained, with full polarisation data
at 1420 MHz.  Observations began in March 1995 and will continue for
approximately 4 years.

\section{Observations}

The observations presented in this work were obtained using the Synthesis
Telescope (ST) of the Dominion Radio Astrophysical Observatory (DRAO), in
1995 March and April.  The array consisted of seven 9m antennas in an 
east-west configuration, observing simultaneously $\lambda$21cm and
$\lambda$74cm continuum and the HI emission line at 1420.406 MHz.  
(The $\lambda$74cm observations, though available, were not used in this
study.)  
The total bandwidth of the $\lambda$21cm maps is 30 MHz.  The HI data 
were acquired with a new 256-channel spectrometer using a bandwidth of 1 MHz.
This corresponds to a channel spacing of 0.824 km s$^{-1}$ with resolution
1.32 km s$^{-1}$.
By tracking a field center for 12 hours and moving the three
mobile antennas between trackings, a complete $u-v$ coverage is obtained
with baselines from 12m to 600m at 4.3m intervals.  

Maps were made by fourier-transforming the edited, calibrated and  gridded
visibilities.  
The $\lambda$21cm maps were then cleaned and self-calibrated.  The HI data
were not cleaned because of the nearly complete $u-v$ coverage and low
sidelobe levels ($<4\%$), and because the dynamic range within each map
($\lesssim$60) was generally 
insufficient to warrant cleaning.  Continuum emission
was subtracted from the HI maps, using an uncleaned continuum map.

Extended structure, corresponding to interferometer spacings less than 12m,
was extracted from single-dish maps.  At $\lambda$21cm, these were obtained
from surveys published by Reich (1982) and Reich \& Reich (1986, 1988).
Single-dish HI data were obtained using the DRAO 26m telescope in July 1995,
and were calibrated by observations of the standard region S7 (Williams 1973).
The ST and single-dish maps were fourier-transformed, filtered and tapered
in a complementary manner, then re-transformed.  The ST maps were then
corrected for the 9m polar diagram, and 
added to the filtered single-dish maps.  In this way
we obtain maps containing full coverage of all structure down to the
resolution of the synthesized maps, which is $\sim 67" \times 59"$ with
major axis oriented at position angle 4\arcdeg east of north.
The RMS noise in the HI maps at the field center is $\sim$2.9 K ($T_B$) per 
channel.  Three fields, with field centers separated equally by $\sim$95',
were conjoined in a mosaic so that the noise is relatively constant over much 
of the image.  The full images will be presented in a later paper, but can be
viewed at http://www.drao.nrc.ca/$\sim$schieven/news\_sep95/ic1396.html.

\section{The Data}

\begin{figure}[h]
\centering
   \leavevmode
   \includegraphics{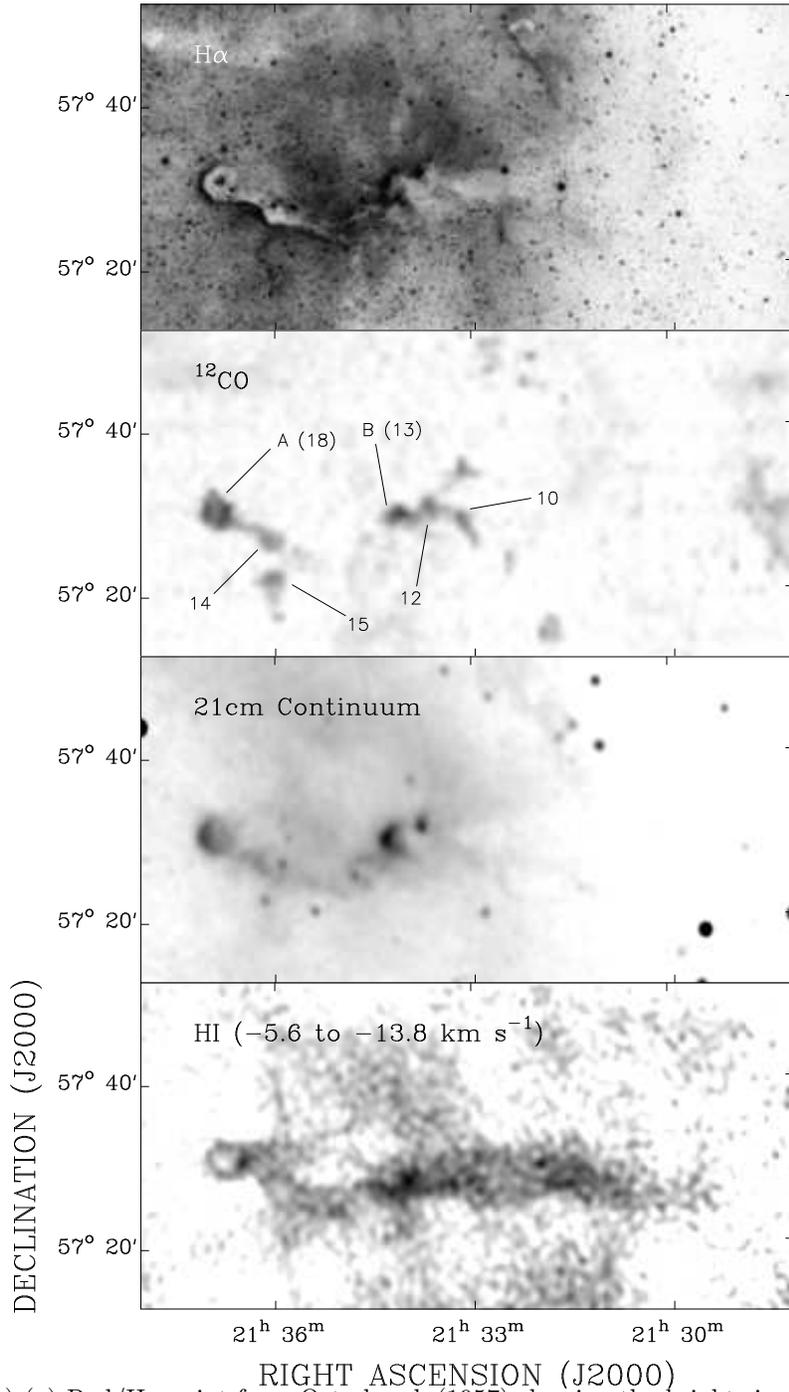}
   \vspace{6.75in}
   \caption{(Plate 1)
(a) Red/H$\alpha$ print 
from Osterbrock (1957) showing the bright rims and dark globules
in the vicinity of globules A and B.  Greyscale in this and in figures below
is linear.  
(b)  Map of $^{12}$CO J=1-0 peak intensity from Patel et al. (1995).  Labels 
refer to rim/globule nomenclature of Pottasch (1956) and Patel et al. (1995).  
(c) Greyscale of $\lambda$21cm continuum emission.  (d)  Atomic hydrogen
emission integrated over velocity interval -5.6 to -13.8 km s$^{-1}$.  
Background emission has been subtracted by removing a twisted plane fitted
to the edges of the displayed box.
}
   \label{Acomb_fig}
\end{figure}

\begin{figure}[h]
\centering
   \leavevmode
   \includegraphics{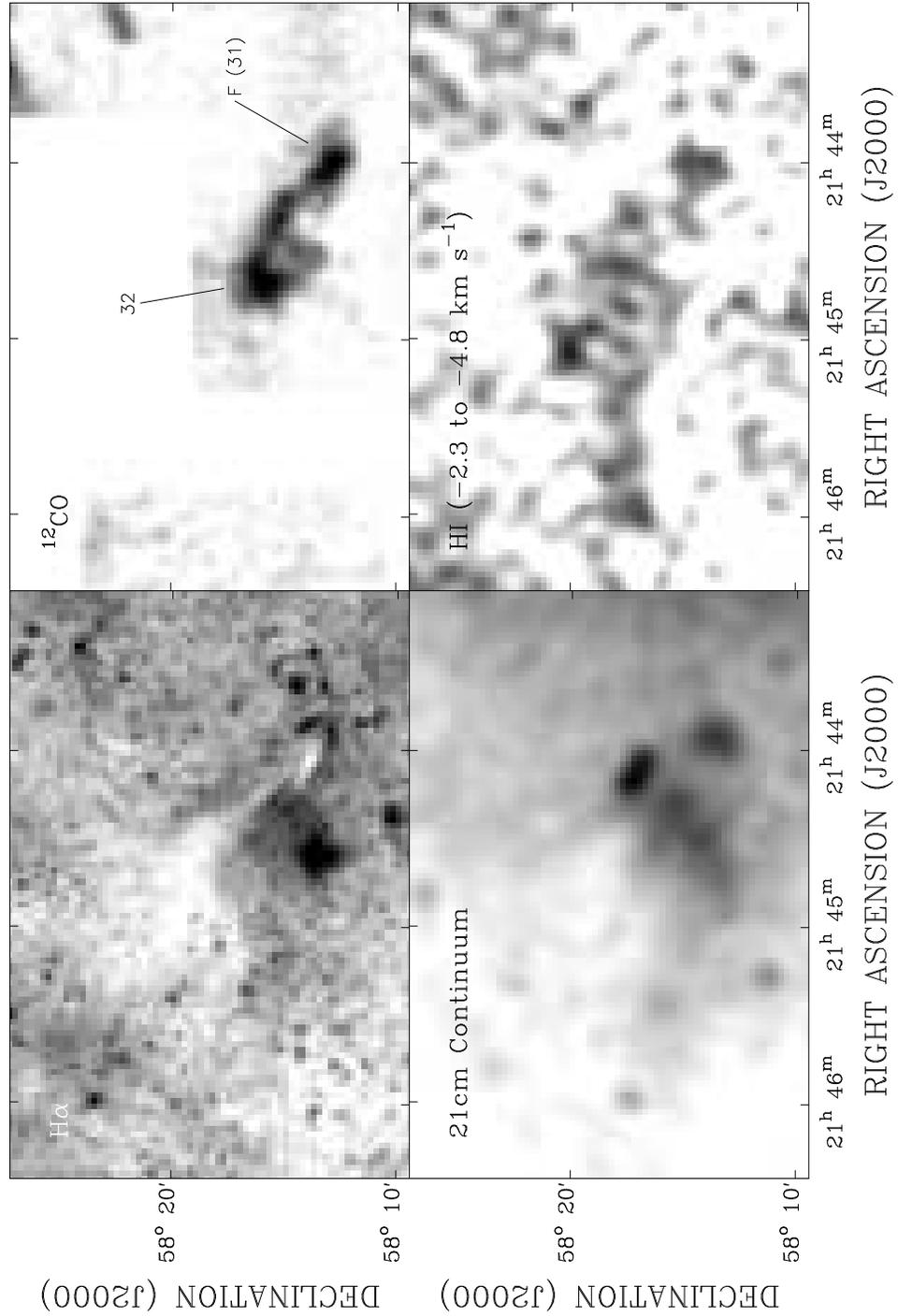}
   \vspace{7.0in}
   \caption{ (Plate 2)
(a) Red print from Osterbrock (1957) showing the bright rims and dark globules
in the vicinity of globule F. 
(b)  Nyquist-sampled 
map of $^{12}$CO J=1-0 peak intensity from Patel et al. (1995).  Labels   
refer to rim/globule nomenclature of Pottasch (1956) and Patel et al. (1995).  
(c) $\lambda$21cm 
continuum emission. The 3K contour of $^{12}$CO is also shown.  
(d)  Atomic hydrogen
emission integrated over velocity interval -2.3 to -4.8 km s$^{-1}$.
Background emission has been subtracted by removing a twisted plane fitted
to the edges of the displayed box.  The 3K contour of $^{12}$CO has been 
overlaid. 
}
   \label{Fcomb_fig}
\end{figure}

While searching for HI associated with the bright rims, we discovered 
remarkable HI features associated with bright 
rims A and B (located 17' and 
37' almost due west of the illuminating star), and with rim F (located
$\sim$1\arcdeg north-east of the illuminating star).  (Nomenclature is from
Pottasch (1956).)    
In Figure \ref{Acomb_fig}a 
(Plate \ref{Acomb_fig}) we show the 
A/B complex of rims/globules from the classic 
red/H$\alpha$ image made by Osterbrock (1957).  Figure \ref{Acomb_fig}b 
is a greyscale map 
of $^{12}$CO J=1-0 emission from Patel et al. (1995), labeled with the 
rim/globule nomenclature of Pottasch (1956) and Patel et al. (1995).
Figures \ref{Acomb_fig}c and d (Plate 
\ref{Acomb_fig}) show the 21cm continuum and 
HI emission respectively from this region.  The HI emission was integrated
over the velocity interval -6 to -14 km s$^{-1}$, and ``background'' emission
was removed by subtracting a twisted plane fitted to a box defined by the
edges of the image.  (The ``background'' emission was subtracted in order to
isolate the HI in the vicinity of the globules.)  
Figure \ref{Fcomb_fig} (Plate \ref{Fcomb_fig}) shows
a similar quartet of images of globule F, including HI integrated over the
velocity interval -2 to +5 km s$^{-1}$.  Despite a careful search, no 
such features were seen toward any
of the other globules.

The HI morphology is clearly related to the globules 
and is quite remarkable.  In the immediate vicinity of globule
A (designated globule 18 by Patel et al. 1995 (see Figure
\ref{Acomb_fig}b, Plate \ref{Acomb_fig})) the HI forms a distinct 
``ring'', with a ``tail'' extending $\sim$10' behind the ring opposite 
the direction of the O6.5 star.
HI also seems to trail behind the ridge of CO extending south-west of 
A (i.e. globule 14 from Patel et al. 1995), and perhaps from the small CO
cloud (globule 15) to the south.  The HI tail from globule 14 extends west
until a bright knot of HI south and west of (i.e. behind) B (globule 13).
The other CO globules in this group (globules 12 and 10) also appear to have
HI associated with them immediately adjacent to the south and west.  Behind
the B complex, another tail of HI extends $\sim$30' due west and ends
beyond the edge of the HII region.  At the
distance of IC1396 (750pc), this implies a projected length of $\sim$6.5pc.

The $\lambda$21cm 
continuum image (Fig. \ref{Acomb_fig}c, Plate \ref{Acomb_fig}) shows bow-shaped
ionization fronts associated with many, but not all, of the globules, and
which closely trace the bright rims.  Also visible is a ``shadow'' west of
globule 12 extending to the edge of the HII region.  Our image is very
similar to the 21cm continuum map of the A and B bright rims made by
Matthews (1979).  (A colour composite of CO, HI and 21cm continuum can be
viewed at http://www.drao.nrc.ca/$\sim$schieven/news\_sep95/ic1396.html).

The $\lambda$21cm image of globule F (Fig. \ref{Fcomb_fig}c; 
Plate \ref{Fcomb_fig}) clearly shows a
bow-shaped ionization front on the leading edge of the globule
(designated 31 in Patel et al. (1995)), as well as a weaker, more diffuse 
ridge along the front edge of globule 32.  (The illuminating star HD206267 
is located $\sim$1\arcdeg approximately south-west of F.)  The HI 
(Fig. \ref{Fcomb_fig}d; Plate \ref{Fcomb_fig}) 
is weaker and less clearly defined than that associated with A 
and B, but its fan-shaped tail is clearly associated with the CO 
(Fig. \ref{Fcomb_fig}b; Plate \ref{Fcomb_fig}).

\begin{figure}[h]
\centering
   \leavevmode
  \includegraphics{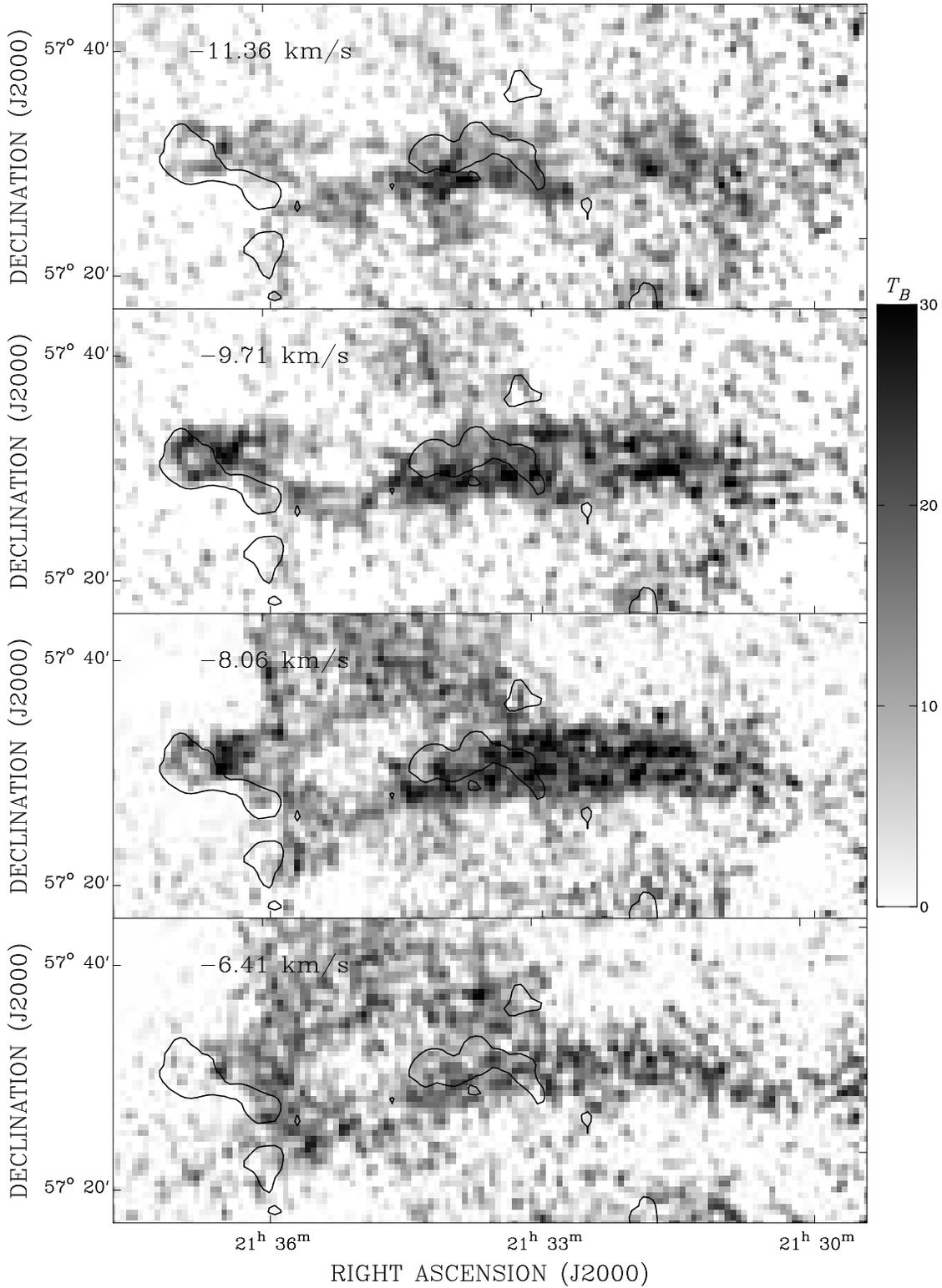}
   \vspace{7.75in}
   \caption{(Plate 3)
Maps of selected channels of HI emission in the vicinity of globules 
A and B.
Background emission has been subtracted by removing a twisted plane fitted
to the edges of the box displayed in Figure \ref{Acomb_fig}.  
The 4K contour of $^{12}$CO has been 
overlaid on each image.  Greyscale intensities are shown by the ``wedge'' at 
right.
}
   \label{Achan_fig}
\end{figure}

\begin{figure}[h]
\centering
   \leavevmode
   \includegraphics{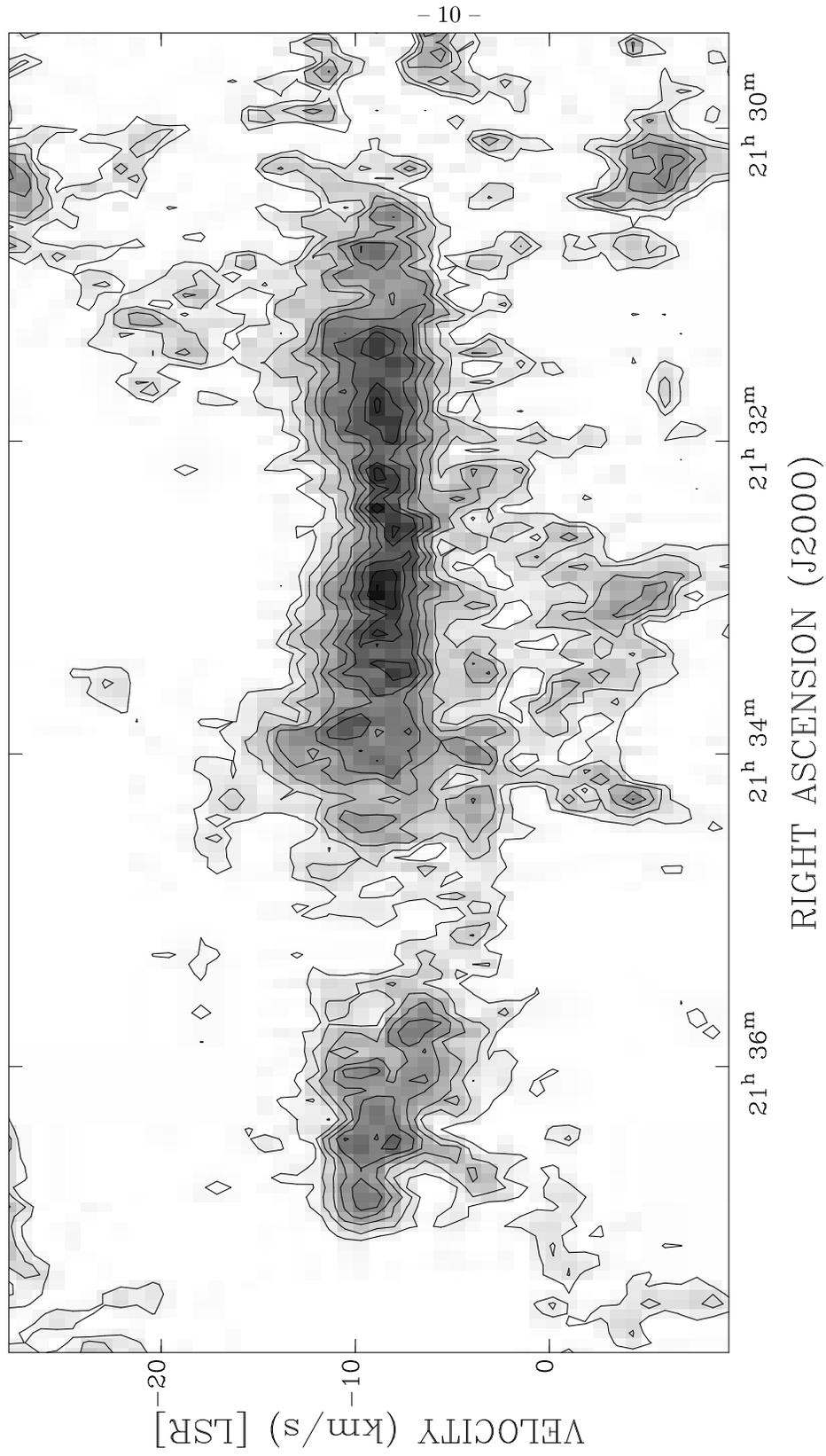}
   \vspace{8in}
   \caption{(Plate 4) 
Spatial-velocity diagram of (background-subtracted) HI along a cut through 
globules A and B, averaging 10 pixels in the declination direction.  Contours
are every 3K (brightness temperature T$_B$).
}
   \label{sv_fig}
\end{figure}

Channel maps at a few velocities of the HI associated with globules A and B 
are shown in Figure \ref{Achan_fig}
(Plate \ref{Achan_fig}).
Over the four channels shown here, which 
represent most of the velocity range of this feature, the ``head'' (i.e.
HI ring in globule A) and ``tail'' (i.e. the rest) show very little 
relative change.  (From -6 to -8 km s$^{-1}$ there is significant 
emission north-west of globule A, but it is unclear if or how this
relates to the rest of the HI.)  In Figure \ref{sv_fig} (Plate \ref{sv_fig}) 
we show a 
spatial-velocity plot, made by averaging a 10-pixel wide ($\sim$6')
row in R.A. centered at declination 57$\arcdeg$30'58".  The center
velocity of the HI is constant through the whole length of the HI feature.
The centroid velocity of the HI associated with globule A is -9.3 km s$^{-1}$
and with B is -9.8 km s$^{-1}$.  (Emission at velocities $\lesssim$-15
km s$^{-1}$ and $\gtrsim$-2 km s$^{-1}$ is not related.)  There does not
appear to be any velocity gradient within the HI, although  
the gas associated with each globule
may be at slightly different velocities.

\begin{figure}[h]
\centering
   \leavevmode
   \includegraphics{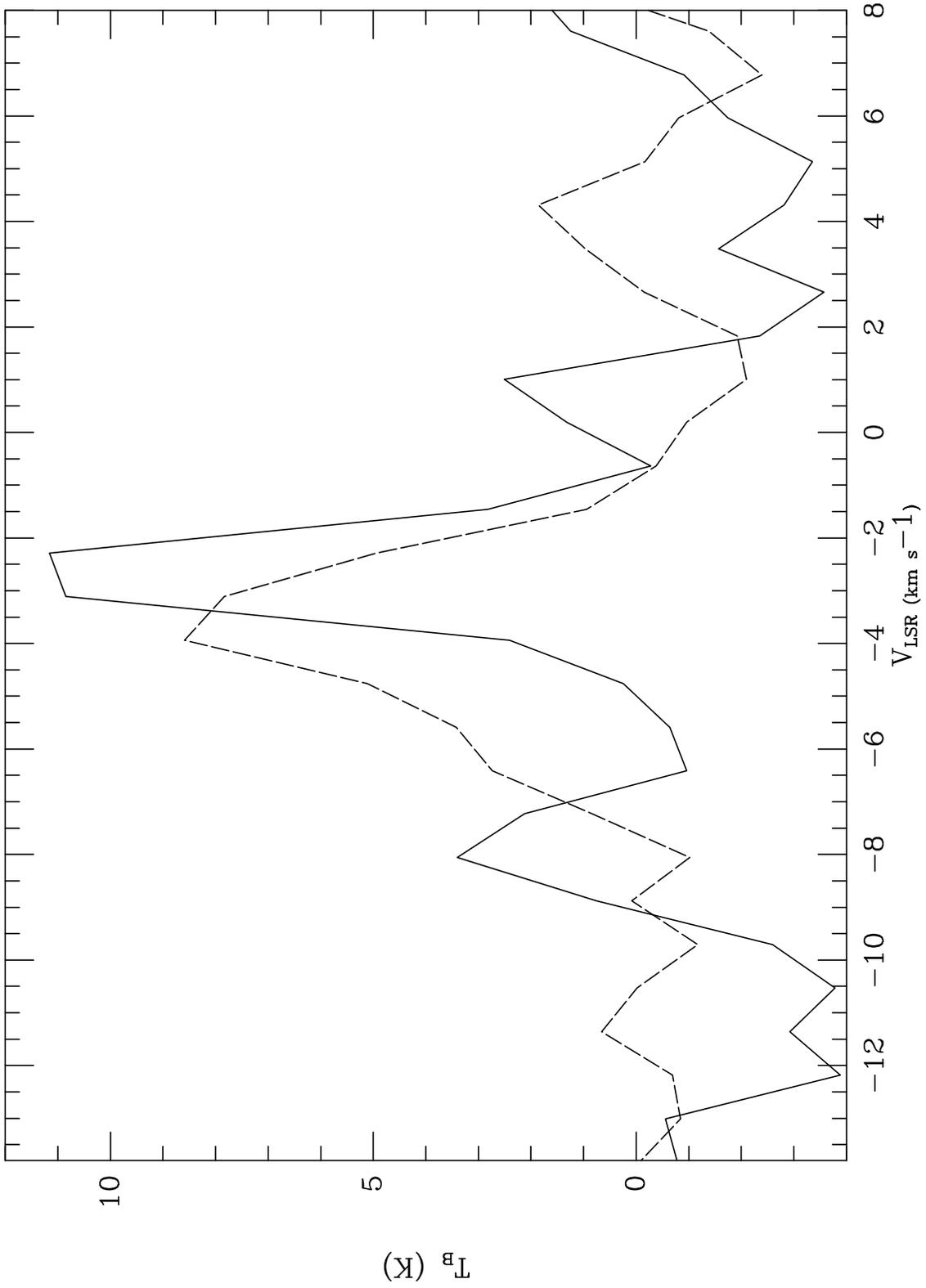}
   \vspace{8in}
   \caption{
Background-subtracted HI spectra in the vicinity of globule F were averaged 
over the HI ``head'' (solid line) and ``tail'' (dashed line), and are shown
here.  Greyscale intensities are shown by the ``wedge'' at right.
}
   \label{Fspec_fig}
\end{figure}

In Figure \ref{Fspec_fig} we display ``background-subtracted'' HI spectra, 
found by averaging HI emission in the immediate vicinity of globule F (31), 
shown as a solid line, and toward the HI ``tail'' of F, 
shown as a dashed line. 
The HI line from the ``tail'' (dashed line) is significantly broader 
(FWHM $\sim$ 3.5 km s$^{-1}$) and blue-shifted (V$_{cen}$ $\sim$ -4.0 
km s$^{-1}$) compared to the ``head'' (solid line, FWHM $\sim$ 2 
km s$^{-1}$, and V$_{cen}$ $\sim$ -2.7 km s$^{-1}$).

\section{Discussion}

The morphology of the HI associated with these globules is remarkable 
and is reminiscent of comets in the solar system.  One possibility 
for the origin of the HI ``tails'' is that, like 
solar system comets, they represent material which has been ablated from the
globule through ionization and/or dissociation and then accelerated away from 
the globule by stellar wind from the central O star.  In this scenario, 
the HI should be moving outward from the globules radially away from the 
central star, at velocities greater than that of the globules.
Patel et al. (1995) found the velocity of the CO
emission from A (and associated globule 14) to range from -7.9 to
-8.2 km s$^{-1}$, while the globules in the vicinity of B (including
10, 12 and 13) range from -4.8 to -5.6 km s$^{-1}$.  The HI associated with
A is thus blue-shifted by $>$1 km s$^{-1}$, and that associated with B
by $>$4 km s$^{-1}$.  The CO velocity of globule F is -3.2 km s$^{-1}$.  
In this case, the HI in the ``head'' is slightly red-shifted compared to 
the CO (by $\sim$ 0.5 km s$^{-1}$), while is the ``tail'' is blue-shifted 
(by $\sim$ 0.8 km s$^{-1}$). 
Patel et al. (1995) found that the CO globules appear to trace
an expanding ellipse.  The radial components 
of the expansion velocities for globules
A, B and F are toward the line-of-sight, so that if the HI is being blown
away from the globules by the stellar wind, the velocity of the HI should be
blue-shifted with respect to
that of the CO, as we observe (except for the ``head'' of F). 
However, in this scenario we might also expect to see an increase in the
velocity of HI with increasing distance from the central star if the
stellar wind were to continue accelerating the gas, as we clearly do
not see for the ``tails'' of A and B, and in addition one might expect the
gas in the accelerated tails to be ionized.

A second possibility
is that the HI ``comets'' are 
ambient material, perhaps predating the HII region, which 
lies within the ``shadow'' of the globules protecting it from ionization 
or acceleration.  In this case, since the globules have been accelerated by
the ``rocket effect'' (Harwit \& Schmid-Burgk 1983) 
while the HI should have been
relatively undisturbed, the HI should be red-shifted with respect to the CO.  
Except for the ``head'' of F, this is not the case.

A third scenario is that the HI is material which, as in the first possibility
above, has been ablated and accelerated from the globules, but the material
has drifted into the shadow of the globules where it is sheltered from 
further ionization or acceleration.  Qualitatively, this scenario seems
the most attractive of the three, since it can account for the blue-shifted 
HI relative to the CO and the lack of acceleration.  We can calculate the mass
of atomic hydrogen in these HI ``comet-tails'' by 
assuming that the optical depth is small.  Then $N_{HI} = 1.823 \times 10^{18}
\int T_B\,dv$ cm$^{-2}$ (Kraus 1982).  
The mass of HI associated with globules A and 
B is then 22 M$_{\odot}$ ($\sim$ 4 M$_{\odot}$ associated with A, 
$\sim$18 M$_{\odot}$ with B), which is $\lesssim$ 5\% of the total mass 
of molecular gas within these globules 
(Patel et al. 1995).  The mass associated with F is 1.5
M$_{\odot}$ ($\sim$0.25 M$_{\odot}$ in the ``head'', $\sim$1.25 M$_{\odot}$
in the ``tail''), which is $<$2\% of the molecular mass.

Is there, however, 
sufficient momentum flux in the stellar wind to have accelerated 
this material by several km s$^{-1}$?  Chlebowski \&
Garmany (1991) have determined the mass loss rate and wind terminal velocity
of HD206267 to be \.{M} = $7 \times 10^{-7}$ M$_{\odot}$ yr$^{-1}$ and
$V_{\infty} = 3.1 \times 10^3$ km s$^{-1}$.  Thus the momentum flux over
$4\pi$ steradians is $\approx 2.1 \times 10^{-3}$ M$_{\odot}$ km s$^{-1}$ 
yr$^{-1}$.  If we assume a 1pc diameter globule, roughly 12.5pc from the
central star (the approximate current radius of the expanding ring (Patel
et al. 1995)), then $\Omega$/4$\pi$ $\approx$ 0.04 and the momentum flux
on the globule is $\approx 8.4 \times 10^{-5}$ M$_{\odot}$ km s$^{-1}$ 
yr$^{-1}$.  The dynamical ages of these tails, neglecting the inclination of
the velocity vector to the line-of-sight, are 1.6 - 2.5 Myr, which is
roughly the age of the HII region.  The material could have been accelerated
for only a short fraction of that time, say $\lesssim$5\% or $\sim 10^5$ yr.
In the immediate vicinity of globule A is $\sim$1 M$_{\odot}$ of atomic gas.
Over $\sim 10^5$ yr, 1 M$_{\odot}$ of material would be accelerated to $\sim$8
km s$^{-1}$.

It is thus plausible that the tails are material which was initially 
ablated and accelerated from the globules, but is now in the shadow of
the dense globules.  There remains the ``head'' of the HI comet associated with
F, which is red-shifted rather than the expected blue-shifted.  According
to the ``rocket effect'' model of Harwit \& Schmid-Burgk (1983), 
it is the action of material being ionized/dissociated on the front surfaces
of the globules which accelerates it away from the star.  The HI ``head'' of
F might be the initially red-shifted ``rocket-exhaust'' before being 
accelerated itself by the stellar wind.  However, we see no such red-shifted
emission on the front surface of globule A (Fig \ref{sv_fig}).

Finally, the ring-like HI structure surrounding globule 18 (rim A) is 
intriguing.  No other globule seems to possess a similar structure.  The 
globule itself is unusual, with a central cavity or hole which can be seen 
both optically (Figure \ref{Acomb_fig}a, Plate \ref{Acomb_fig}) and in 
molecular emission (Wooten et al 1983; Nakano et al. 1989; Patel et al. 
1995).  Inside the cavity are two stars, LkH$\alpha$ 349 and 
LkH$\alpha$ 349/c, which are young stellar objects, the former 
of which may be on its way to becoming a Herbig Be star 
(Hessman et al. 1995).  These stars are unlikely to have 
ionized or dissociated the gas in the cavity.  Instead, the 
cavity was likely evacuated during an earlier outflow stage 
of one or both of these stars (Nakano et al. 1989).

Near the outside edge of the ``backside'' of the globule, $\sim$100'' 
south-west of LkH$\alpha$ 349, is a B3V star VDB 142 (HD 239710, AG+57 1457).  
Optical images of the region (Osterbrock 1957; P. Boltwood, private 
communication) show that this star is surrounded by diffuse nebulosity, 
suggesting that this star is physically associated with the globule.  
A B3V star can create a small HII region and a larger HI photodissociation 
region (Roger \& Dewdney 1992).  There is some indication of an enhancement 
of HI intensity toward this star.  Thus the ring morphology of this globule 
may the result of ionization and photodissociation on both the front and 
back surfaces of the globule, plus the evacuation of the central cavity.

\section{Summary}

We have mapped the $\lambda$21cm 
continuum and HI emission at $\sim$1' resolution,
covering the nearby (750pc) HII region IC1396, which contains a number
of cometary globules with bright, ionized rims.  
A small number of globules have long comet-like ``tails'' of HI extending as 
much as 30' (8pc), pointing away from the central ionizing star (HD 206267).  
The masses
of these HI structures range 
from $\sim$4 M$_{\odot}$ to $\sim$20 M$_{\odot}$, which is a small fraction
($<$5\%) of the molecular mass of the globules.
The HI is blue-shifted in velocity relative to the CO, as would be expected
if the HI tails were material being accelerated away from the globules.
However, there is little evidence for acceleration within the HI tails.
There is sufficient momentum flux in the stellar wind originating from the 
central star to have accelerated this much material in a relatively 
short time.  
These observations are thus 
consistent with a scenario in which the globules are
being ionized and photodissociated 
on the front surface by the central star, and
the ablated material is then being blown away from the globule by the stellar
wind after which it drifts into the shadow of the globule, 
to form the long comet-like tails of atomic hydrogen.  An intriguing 
ring-like structure surrounding one globule is likely caused by 
ionization/dissociation of the front and back surfaces of the globule by
two different stars.

This work represents the first results of the Galactic Plane Survey now
underway at the Dominion Radio Astrophysical Observatory.

\acknowledgements

G.M.-S. was 
supported at DRAO by a Research Associateship from the National Research
Council of Canada.  
T.X. 
is supported in part by the NSF grant AST9314847, and he 
is grateful to the faculty at the Laboratory for Millimeter-wave Astronomy
for creating a superb research environment and for granting the Frank Kerr
fellowship to him.  He further acknowledges useful discussions with Paul 
Goldsmth and Leo Blitz in the planning stages of the IC1396 project.
The Galactic Plane Survey is funded by a Collaborative Special Projects grant
from the National Sciences and Engineering Research Council of Canada.

\end{document}